\documentclass[mathleft
]{an}
\usepackage{graphicx}

\usepackage{times}
\begin{document}

\Pagespan{1}{}
\Yearpublication{2006}%
\Yearsubmission{2005}%
\Month{11}%
\Volume{999}%
\Issue{88}%

\title{The exceptional Herbig Ae star HD\,101412:\\The first detection of resolved magnetically split 
lines and the presence of chemical spots in a Herbig star\thanks
{Based on observations obtained at the European Southern Observatory (ESO programmes 077.C-0521(A) and 383.C-0684(A)).}}

\author{{S. Hubrig\inst{1}\fnmsep\thanks{Corresponding author: \email{shubrig@aip.de}}}
\and
M. Sch\"oller\inst{2}
\and
I. Savanov\inst{3}
\and
J.~F.~Gonz\'alez\inst{4}
\and
C.~R.~Cowley\inst{5}
\and
O.~Sch\"utz\inst{6}
\and
R.~Arlt\inst{1}
\and
G.~R\"udiger\inst{1}}

\titlerunning{The exceptional Herbig Ae star HD\,101412}
\authorrunning{S. Hubrig et al.}
\institute{
Astrophysikalisches Institut Potsdam, An der Sternwarte 16, 14482 Potsdam, Germany
\and
European Southern Observatory, Karl-Schwarzschild-Str.\ 2, 85748 Garching bei M\"unchen, Germany
\and
Institute of Astronomy, Russian Academy of Sciences, Pyatnitskaya 48, Moscow 119017, Russia
\and
Instituto de Ciencias Astronomicas, de la Tierra, y del Espacio (ICATE), 5400 San Juan, Argentina
\and
Department of Astronomy, University of Michigan, Ann Arbor, MI 48109-1042, USA
\and
European Southern Observatory, Alonso de Cordova 3107, Casilla 19001, Santiago 19, Chile
}



\abstract{
In our previous search for magnetic fields in Herbig~Ae stars, we pointed 
out that HD\,101412 possesses
the strongest magnetic field among the Herbig Ae stars and
hence is of special interest for follow-up studies of magnetism among young 
pre-main-sequence stars.
We obtained high-resolution, high signal-to-noise UVES and a few lower 
quality HARPS spectra revealing the presence of
resolved magnetically split lines.
HD\,101412 is the first Herbig Ae star for which the rotational
Doppler effect was found to be small in comparison to the magnetic splitting 
and  several spectral lines
observed in unpolarized light at high dispersion are resolved into 
magnetically split components.
The measured mean magnetic field modulus varies from 2.5 to 3.5\,kG, while the mean
quadratic field was found to vary in the range of 3.5 to 4.8\,kG.
To determine the period of variations,
we used radial velocity, equivalent width, line width, and line asymmetry measurements of variable spectral
lines of several elements, as well as magnetic field measurements.
The period determination was done using the Lomb-Scargle method.
The most pronounced variability was detected for spectral lines of 
\ion{He}{i} and the iron peak elements,
whereas the spectral lines of CNO elements are only slightly variable. From 
spectral variations and
magnetic field measurements we derived a potential rotation period P$_{\rm 
rot}$=13.86\,d, which has to
be proven in future studies with a larger number of observations.
It is the first time that the presence of element spots is detected on the 
surface of a Herbig Ae/Be star. Our previous study of Herbig Ae stars revealed a trend towards 
stronger magnetic fields for younger Herbig Ae stars, confirmed by statistical tests. This is in 
contrast to a few other (non-statistical) studies claiming that magnetic Herbig Ae stars are progenitors of 
the magnetic Ap stars. New developments in MHD theory show that the measured  magnetic field 
strengths are compatible with a current-driven
instability of toroidal fields generated by differential rotation
in the stellar interior. This explanation for magnetic intermediate-mass 
stars could be an alternative to a frozen-in fossil field.
}

\maketitle

\section{Introduction}

It is generally accepted that accretion from a disk is an integral
phase of star formation. 
A number of Herbig Ae stars and classical T\,Tauri stars are surrounded 
by active accretion disks and, probably, most of the excess emission seen at
various wavelength regions can be attributed to the interaction of the disk
with a magnetically active star (e.g.\ Muzerolle et al.\ \cite{Muzerolle2004}).
This interaction is generally referred to as
magnetospheric accretion. Recent magnetospheric accretion models for these stars 
assume a dipolar magnetic field
geometry and accreting gas from a circumstellar disk falling ballistically 
along the field lines onto the stellar surface. 

In our recent study, we reported new detections of a magnetic field at a level higher 
than 3$\sigma$ in six Herbig Ae 
stars (Hubrig et al.\ \cite{hubrig09}).
In that work, the largest longitudinal magnetic field, $\left<B_{\rm z}\right>$\,=\,$-$454$\pm$42\,G, was 
detected in the Herbig Ae star HD\,101412 using hydrogen lines. The presence of a magnetic field in this star 
was already reported by Wade et al.\ (\cite{Wade2005}), who suggested an age of $\sim$2\,Myr.
HD\,101412 was observed by us on two consecutive nights
in May 2008, revealing a change of the mean longitudinal magnetic field strength by $\sim$100\,G, from  
$\left<B_{\rm z}\right>$\,=\,$-$454$\pm$42\,G to  $\left<B_{\rm z}\right>$\,=\,$-$317$\pm$35\,G.
As we already reported in this previous study, UVES spectra retrieved from the ESO archive 
exhibited a conspicuous 
variability of metal lines reminiscent of peculiar 
main-sequence magnetic Ap stars. Since the age of HD\,101412 is only $\sim$2\,Myr 
this star is observed at an evolutionary stage where the magnetic field plays a critical role in controlling 
accretion and stellar wind.  Hence, the magnetic nature of this object makes it a prime 
candidate for studies of the relation 
between magnetic field and physical processes occurring during stellar formation.
In this work we present a more detailed study of the magnetic field and spectral variability 
of this unique Herbig Ae star 
based on high-resolution high signal-to-noise UVES spectra and a few lower quality HARPS spectra 
acquired during 2009.

\section{Observations and measurements}
\subsection{Spectroscopic material}

Since this star exhibits the strongest magnetic field and hence is of special interest,
we applied for observing time with UVES at Kueyen/UT2 at the VLT to obtain multi-epoch high-resolution 
spectra in service mode to study the magnetic field, to characterise the behaviour of the chemical elements 
on the surface of HD\,101412, and to determine the rotation period.
Due to the replacement of the UVES red mosaics MIT CCD by another chip with an improved quantum efficiency
in the far red optical spectral region in May-July 2009, instead of the requested 12 UVES spectra it was  only
possible to obtain five UVES spectra. In these observations we used the UVES DIC2 390+760 standard setting covering
the spectral range from 3290\,\AA{} to 4500\,\AA{} in the 
blue arm and the spectral range  from  5680\,\AA{} to 9460\,\AA {} 
in the red arm. The slit width was set to $0\farcs{}3$ for the red arm
and $0\farcs{}4$ for the blue arm, corresponding to a resolving power 
of $\lambda{}/\Delta{}\lambda{} \approx 110,000$ and 
$\approx 90,000$, respectively.
The spectra have been reduced with the software packages in the MIDAS environment provided by ESO
to extract one-dimensional spectra. 
Four additional spectra were taken during technical tests with the HARPS spectrograph installed at the 3.6\,m telescope
on La Silla on June 3 2009 and July 4 2009. The spectra cover the spectral range  
from  3780\,\AA{} to 6860\,\AA {}. To enlarge our material 
we also use in this study high-resolution UVES spectra with the standard setting RED580 
in the spectral range 4780\,\AA{} to 6808\,\AA {} available from the ESO archive. These spectra were taken at 
a resolution of $\sim$80,000.

\begin{table*}
\caption[]{
Logbook of the spectroscopic observations of HD\,101412, and of the
magnetic field measurements.}
\begin{center}
\begin{tabular}{lccccc}
\hline \hline\\[-7pt]
\multicolumn{1}{c}{Instrument} &
\multicolumn{1}{c}{MJD} &
\multicolumn{1}{c}{S/N} &
\multicolumn{1}{c}{RV} &
\multicolumn{1}{c}{$\left<B\right>$} &
\multicolumn{1}{c}{$\langle B_q\rangle$}\\
\multicolumn{1}{c}{ } &
\multicolumn{1}{c}{ } &
\multicolumn{1}{c}{ } &
\multicolumn{1}{c}{[km\,s$^{-1}$]} &
\multicolumn{1}{c}{[kG]} &
\multicolumn{1}{c}{[kG]}\\
\hline\\[-7pt]
UVES\,RED~580      & 53870.986 & 180 & 17.28 & 3.52 & 4.85 \\
UVES\,RED~580      & 53872.041 & 220 & 16.72 & 2.97 & 4.25 \\
UVES\,RED~580      & 53919.996 & 150 & 15.79 & 2.60 & 3.81 \\
UVES\,DIC2~390+760 & 54928.120 & 340 & 16.40 & 2.65 & 3.76 \\
UVES\,DIC2~390+760 & 54930.076 & 320 & 16.36 & 2.53 & 3.55 \\
UVES\,DIC2~390+760 & 54936.153 & 310 & 17.37 & 2.51 & 2.90 \\
UVES\,DIC2~390+760 & 54943.151 & 305 & 16.11 & 2.73 & 3.53 \\
UVES\,DIC2~390+760 & 54951.065 & 315 & 17.06 & 2.87 & 3.52 \\
HARPS              & 54985.116 & 75  & 16.56 & 2.54 & --   \\
HARPS              & 54985.131 & 80  & 16.46 & 2.58 & --   \\
HARPS              & 55017.000 & 90  & 16.96 & 2.79 & --   \\
HARPS              & 55017.018 & 85  & 16.78 & 2.75 & --   \\
\hline
\end{tabular}
\end{center}
\label{tab:log_meas}
\end{table*}

The logbook of the available spectroscopic observations is presented in Table~\ref{tab:log_meas}.
In the first column we indicate the spectrograph and the setting used. The MJD values for the middle of each exposure
are listed in Column~2 and in Column~3 we present achieved signal-to-noise ratios (S/N) in the 
spectral region around 6150\,\AA {}. The radial velocities listed in Column~4 were measured using numerous Fe lines. 
The measurement accuracy is of the order of 0.1\,km\,s$^{-1}$ for UVES spectra and 0.3\,km\,s$^{-1}$ for HARPS spectra.

\begin{figure}
\centering
\includegraphics[width=0.45\textwidth]{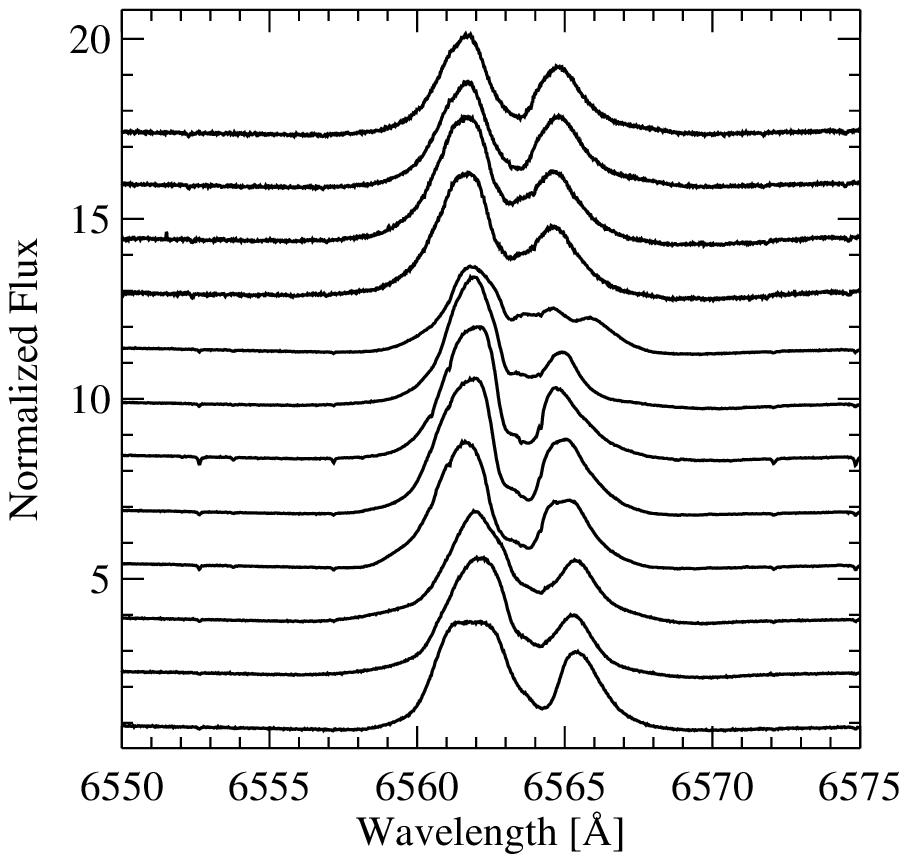}
\includegraphics[width=0.45\textwidth]{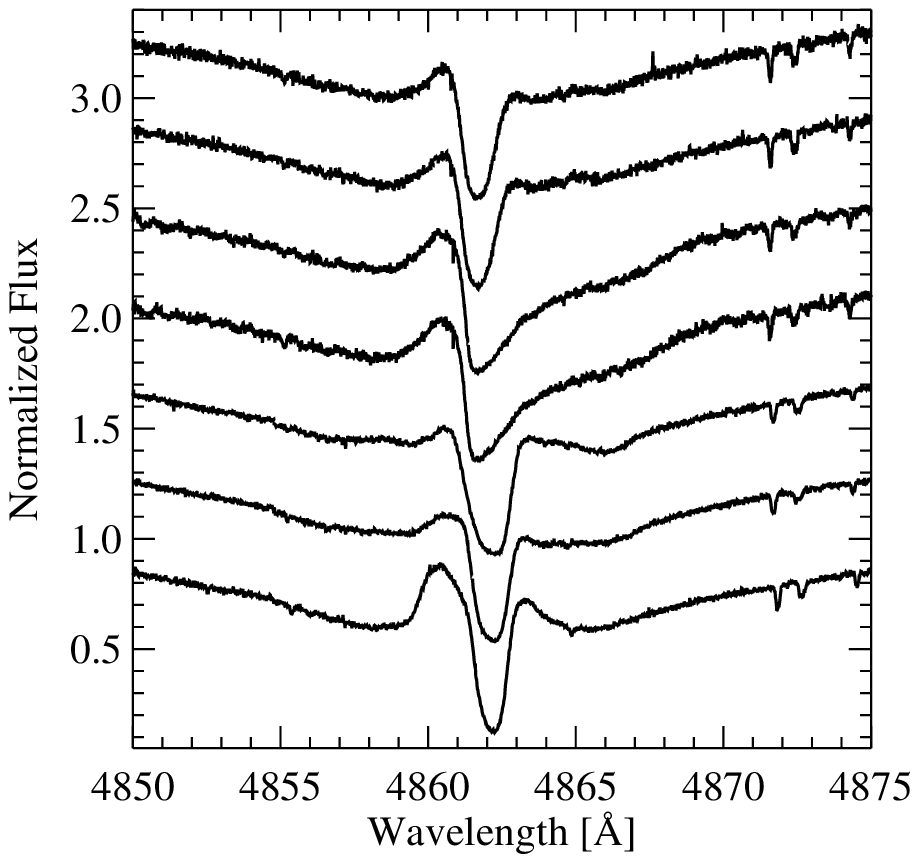}
\caption{
UVES and HARPS spectra of HD\,101412 obtained at different epochs in spectral regions 
around the H$\alpha$ line (top) and the H$\beta$ line (bottom). The spectra are presented with MJD dates increasing 
from bottom to top and offset in vertical direction for clarity.
Please note that H$\beta$ falls into the gap of the UVES\,DIC2~390+760 setting.
}
\label{fig:3}
\end{figure}

In Fig.~\ref{fig:3} we present the most prominent emission features appearing 
in the H$\alpha$ and H$\beta$ lines.
Double-peaked H$\alpha$ emission lines indicate the presence of a temporal variability of the circumstellar 
disk. Temporal variability is also clearly detectable in the H$\beta$ and other 
Balmer lines. van der Plas et al.\ (\cite{plas08}) suggest that the disk of HD\,101412
is in transition between flaring and being self-shadowed.

 \begin{figure}
\centering
\includegraphics[width=0.45\textwidth]{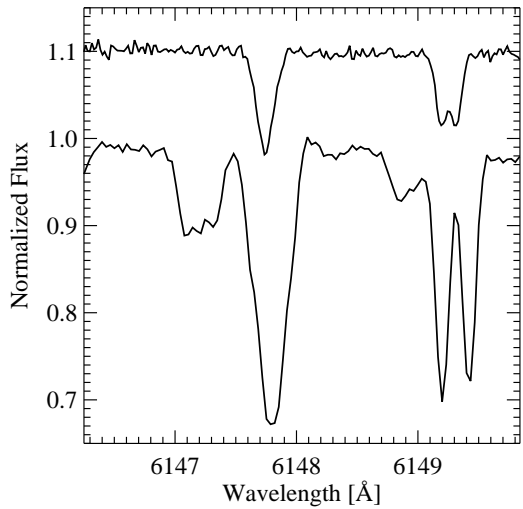}
\caption{
Portion of the spectra of HD\,101412 (vertically shifted in intensity by 0.1 for clarity) and of the typical Ap star HD\,116458 
containing the lines of \ion{Fe}{ii} $\lambda$~6147.741 and $\lambda$~6149.258. 
}
\label{fig:5}
\end{figure}

\subsection{The mean magnetic field modulus}

Our previous inspection of three UVES spectra retrieved from the ESO archive, which were recorded on three different dates,
indicated distinct variations of line intensities and spectral profiles. Specifically, the 
Zeeman doublet \ion{Fe}{ii} 
at $\lambda$~6149.258, which is the best diagnostic line to detect surface magnetic fields in slowly 
rotating classical Ap stars with strong magnetic fields, appeared slightly resolved, indicating 
the presence of a rather strong surface magnetic field. The study of this Zeeman doublet in 
magnetic stars presents an excellent opportunity to 
determine in a straightforward, mostly approximation-free, model-independent way, and
with particularly high precision the mean magnetic field modulus $\left<B\right>$, that is, 
the average over the visible stellar hemisphere of the modulus of the magnetic vector, weighted by the 
local line intensity. The newly acquired UVES spectra confirm our previous suspicion of the presence of 
magnetically split Zeeman patterns. In Fig.~\ref{fig:5} we present the  \ion{Fe}{ii} 
$\lambda$~6149.258 line exhibiting a resolved Zeeman doublet structure at MJD 54936.153. For comparison, we 
present in the lower panel of this figure a typical example for splitting in the spectrum of the bright 
A0p star HD\,116458, 
with a magnetic field modulus of the order of 4.7\,kG. 

\begin{figure}
\centering
\includegraphics[width=0.45\textwidth]{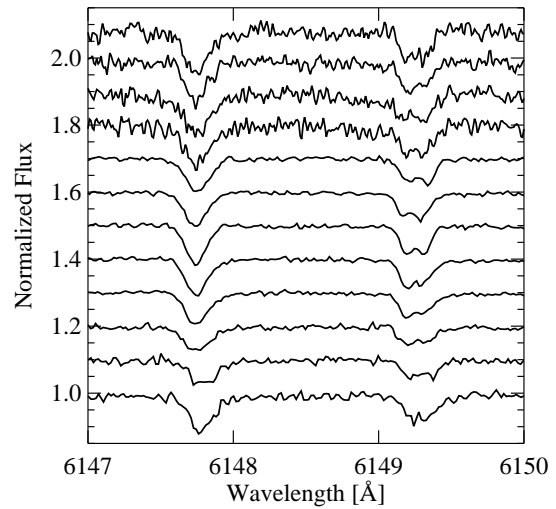}
\caption{
Line profile variations of the Zeeman doublet \ion{Fe}{ii} $\lambda$~6149.258 in UVES and HARPS spectra obtained at different
epochs. The spectra are presented with MJD dates increasing from bottom to top.  
}
\label{fig:1}
\end{figure}

In the approximation of the linear Zeeman effect, the mean magnetic field modulus is
related to the wavelength separation of the Zeeman components through the relation 
\begin{displaymath}
\centering
\left<B\right>=\Delta\lambda/(9.34\cdot10^{-13}\;\lambda_{c}^{2}\;g_{\rm eff}),
\end{displaymath}
where $\left<B\right>$ is the mean magnetic field modulus in Gauss,
$\lambda_{\rm c}$ is the central wavelength of the line in \AA{}, 
${\rm \Delta}\lambda$ is the wavelength separation between the 
centroids of the $\sigma$-components and 
$g_{\rm eff}$ is the effective  Land\'e  factor. 
The wavelengths ${\rm \Delta}\lambda$ of the centres of gravity of the
split doublet components are usually determined either by direct integration of the whole component profiles 
or by fitting a Gaussian simultaneously to each of them (see Mathys et al.\ \cite{mathys97} for more details).
The multi-Gaussian fitting is preferred in our measurements since the $\lambda$~6149.258 line is 
only marginally resolved. 
The typical standard deviation of our measurements is of the order of $\approx$30--50\,G obtained 
for the spectra
in which the Zeeman components are well resolved and  mostly symmetric.
In the worst cases, using the spectra with low S/N, the accuracy of the measurement is of the
order of 100--200\,G. The results of our measurements are presented in Column~5 of Table~\ref{tab:log_meas}.
The observed variable asymmetry of the split components
is usually explained by the variable combination of Zeeman and Doppler effects across the stellar surface.
The behaviour of the red and blue split components of the \ion{Fe}{ii} $\lambda$~6149.258 line 
in all available spectra of HD\,101412 is presented in Fig.~\ref{fig:1}. In the same Figure the
neighbouring line \ion{Fe}{ii} $\lambda$~6147.741, which is a Zeeman pseudo-quadruplet, appears variable too.
At the epoch MJD54936.153 the shape of \ion{Fe}{ii} $\lambda$~6147.741 is clearly triangular indicating that 
the $\sigma$-components of the line are very weak at this rotation phase due to the predominance of a transversal field. 

\begin{figure}
\centering
\includegraphics[width=0.45\textwidth]{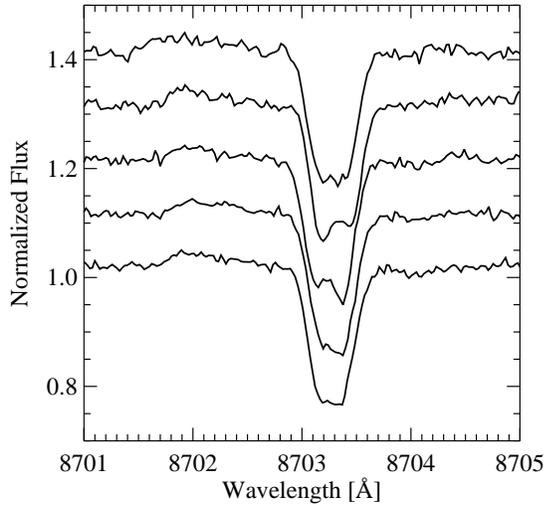}
\caption{
Variations of the observed profile of the Zeeman doublet \ion{N}{i} $\lambda$~8703.24 in the five UVES 
spectra obtained in the near-IR spectral region.
The spectra are presented with MJD dates increasing from bottom to top.  
}
\label{fig:NI}
\end{figure}
An additional example of a Zeeman split doublet is presented in Fig.~\ref{fig:NI}.
The line \ion{N}{i} $\lambda$~8703.24 arises from a transition between two levels having a total angular momentum 
quantum number $J$=1/2, of which one has a Land\'e factor equal to zero 
(that is, this level is unsplit in a magnetic field). 
Partial splitting was also detected in \ion{Fe}{i} lines at $\lambda$~4259.997 and $\lambda$~6336.82, and
in the \ion{Fe}{ii} line $\lambda$~6238.392.
The field modulus measured in {the spectral lines} with the most
evident splitting is in good agreement with our measurements using the \ion{Fe}{ii} $\lambda$~6149.258 line.

\begin{figure}
\centering
\includegraphics[width=0.45\textwidth]{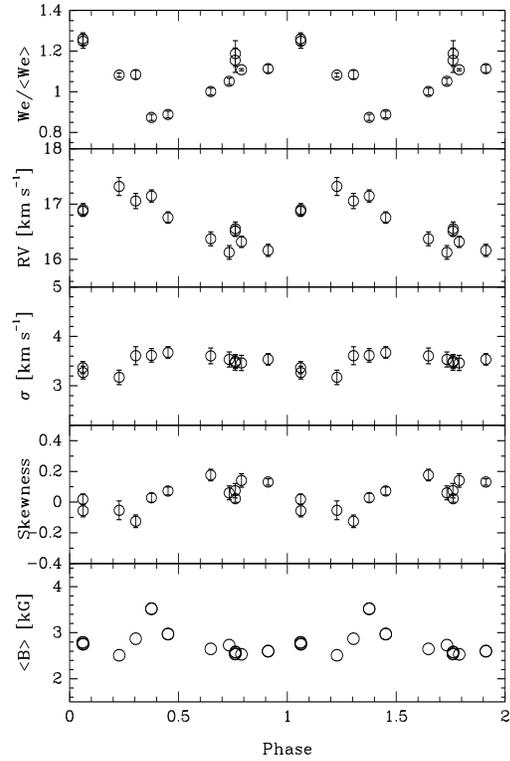}
\caption{
Variations of equivalent width, radial velocity, line width, line asymmetry, and 
mean magnetic field modulus as a function of the rotational phase.
}
\label{fig:per}
\end{figure}

\subsection{The mean quadratic magnetic field}

Another approach to study the presence of magnetic fields 
is to determine the value of the mean quadratic magnetic field,
\begin{displaymath}
\langle B_q\rangle= (\langle B^2\rangle + \langle B_z^2\rangle)^{1/2},
\end{displaymath}
which is derived through the application of the moment technique, 
described in detail by Mathys (\cite{mathys95}) and Mathys \& Hubrig (\cite{mathys06}). Here, $\langle B^2\rangle$ is 
the mean square magnetic field modulus (the average over the stellar 
disk of the square of the modulus of the field vector, weighted by 
the local emergent line intensity), while $\langle B_z^2\rangle$ is 
the mean square longitudinal field (the average over the stellar 
disk of the square of the line-of-sight component of the magnetic 
vector, weighted by the local emergent line intensity). 
The mean quadratic magnetic field is determined from the study of the 
second-order moments of the line profiles recorded in unpolarised 
light (that is, in the Stokes parameter $I$). The analysis is usually based on a consideration of samples 
of reasonably unblended lines of \ion{Fe}{i} and \ion{Fe}{ii} in spectra with a rather high S/N. Our measurements
using exclusively UVES spectra, which have higher S/N compared to HARPS spectra, are presented in the last column of 
Table~\ref{tab:log_meas}. 
The accuracy of measurements of $\langle B_q\rangle$ is usually less than for the measurements of the mean magnetic 
field modulus, accounting for 0.3--0.6\,kG in our study. Interestingly, the lowest quadratic field $\langle B_q\rangle$=2.9\,kG
was measured at the epoch MJD54936.153, where the shape of \ion{Fe}{ii} $\lambda$~6147.741 appears triangular (Fig.~\ref{fig:1},
sixth spectrum from the bottom).
This low value is probably due to a smaller contribution of the mean longitudinal magnetic field.
Clearly, to properly constrain the magnetic field geometry,
additional spectropolarimetric measurements including the measurement of the longitudinal magnetic field 
to better sample the rotation phases of HD\,101412, are necessary. 

\subsection{Spectrum variability}

{The inspection of our spectroscopic material indicates the 
presence of the elements He, C, N, O, Na, Mg, Al, Si, S, 
Ca, Sc, Ti, V, Cr, Mn, Fe, Co, Ni, Zn, Sr, Y, Zr, and Ba. 
Almost all spectral lines show variations in line intensity 
and line profile, with the most pronounced variability 
detected for lines of the elements
He, Si, Mg, Ca, Ti, Cr, Fe Sr, Y, Zr, and Ba. Since also the magnetically insensitive lines show clear profile variations,
we assume that the detected spectral variability is a combination of both Zeeman splitting and abundance spots. 
The potential period of variations, which we assume to correspond to the rotation period, analogous to magnetic Ap stars, 
was determined from line profile variations and the magnetic field modulus measurements.

The line profiles of the Fe lines have been characterized  
calculating the first statistical moments of the flux distribution
within the line profiles.
Twenty Fe lines have been measured, from which the mean value and its
standard error were calculated.
A probable rotation period of P$_{\rm rot}$=13.86\,d was derived from the
measurements of radial velocity, equivalent width, and magnetic field modulus,
using the Lomb-Scargle method (Press \& Rybicki \cite{Press89}).
In Fig.~\ref{fig:per} we present the variations of equivalent width, 
radial velocity, line width, line asymmetry, and 
mean magnetic field modulus as a function of the rotational phase. 
To combine the equivalent widths for different spectral lines,
the measurements of the individual lines were normalised with the 
mean value for each spectral line. Thus, the upper panel shows the typical 
fractional variation in the equivalent width of the Fe lines.

The variations of the measured parameters are significant and are correlated 
with each other, as it is expected for the presence of 
abundance patches on the surface: for example,
there is a 1/4 cycle shift between the equivalent width curve and the radial velocity curve.
At phase zero, when the equivalent width of the Fe lines is at maximum and
the radial velocity in increasing, the
region with the largest Fe abundance would be facing the observer.
The maximum of the surface magnetic field corresponds to the minimum of the Fe abundance.
We note that the number of spectra is rather small and further observations 
are needed to confirm this periodicity.}

\begin{figure*}
\centering
\includegraphics[width=0.3\textwidth]{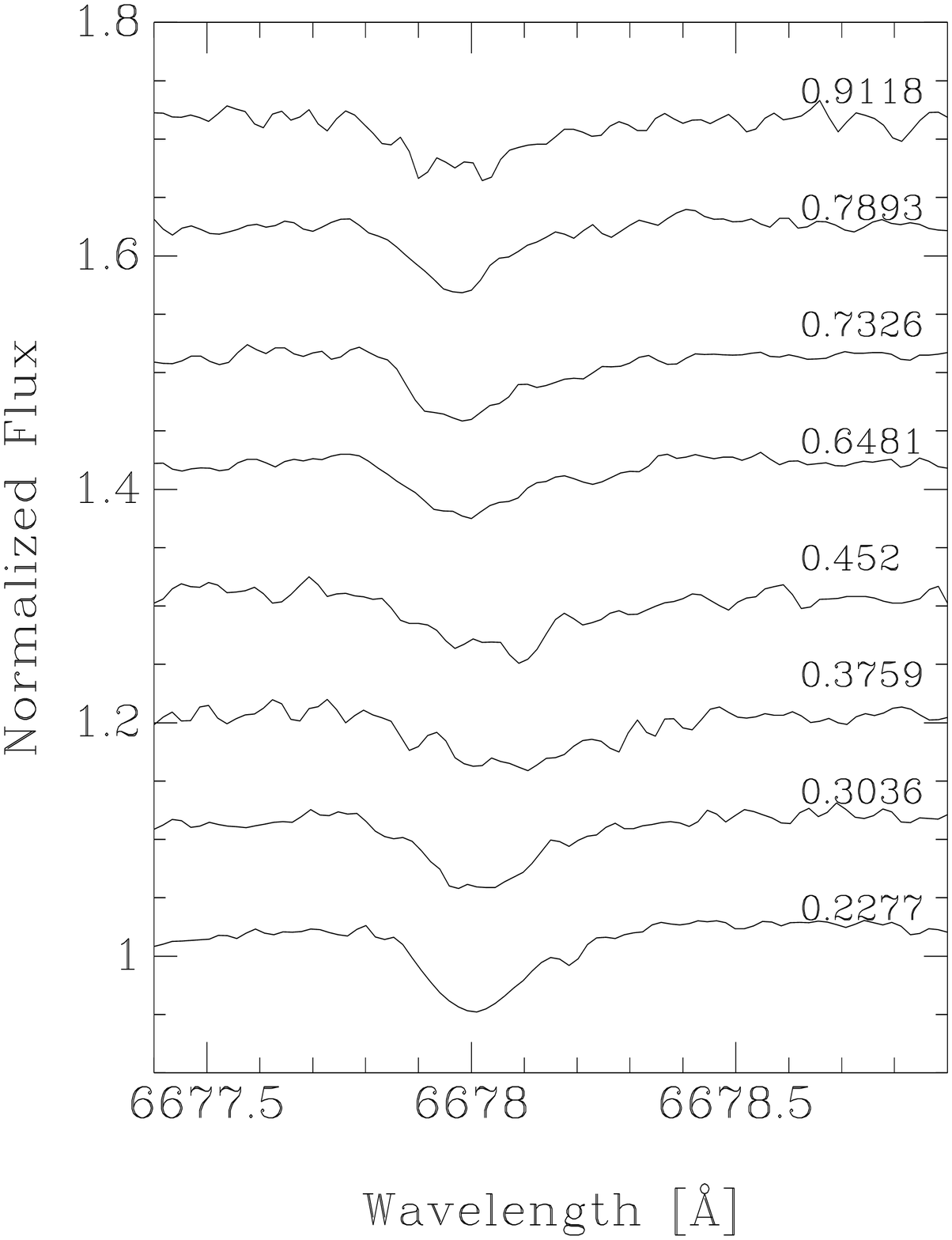}
\includegraphics[width=0.3\textwidth]{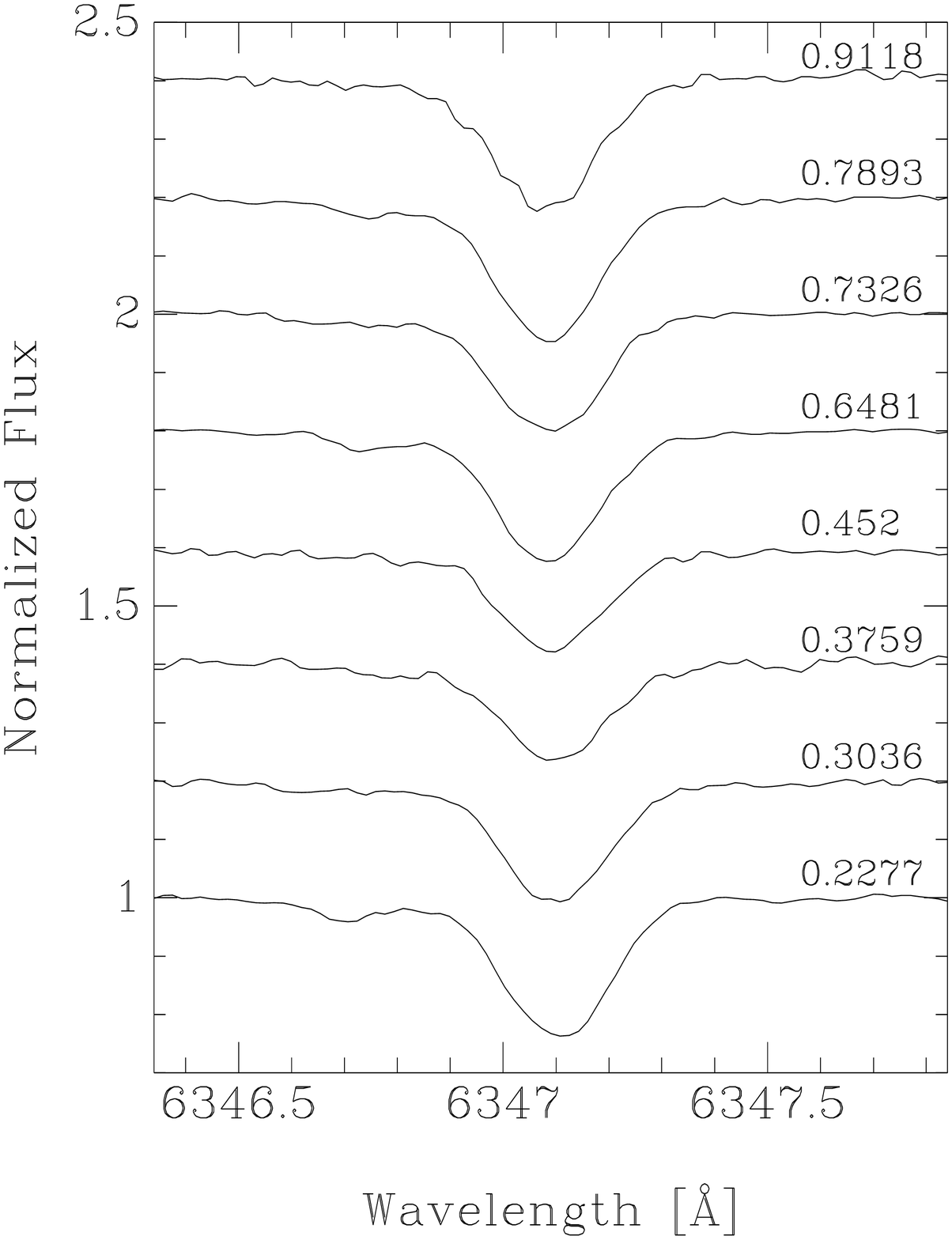}
\includegraphics[width=0.3\textwidth]{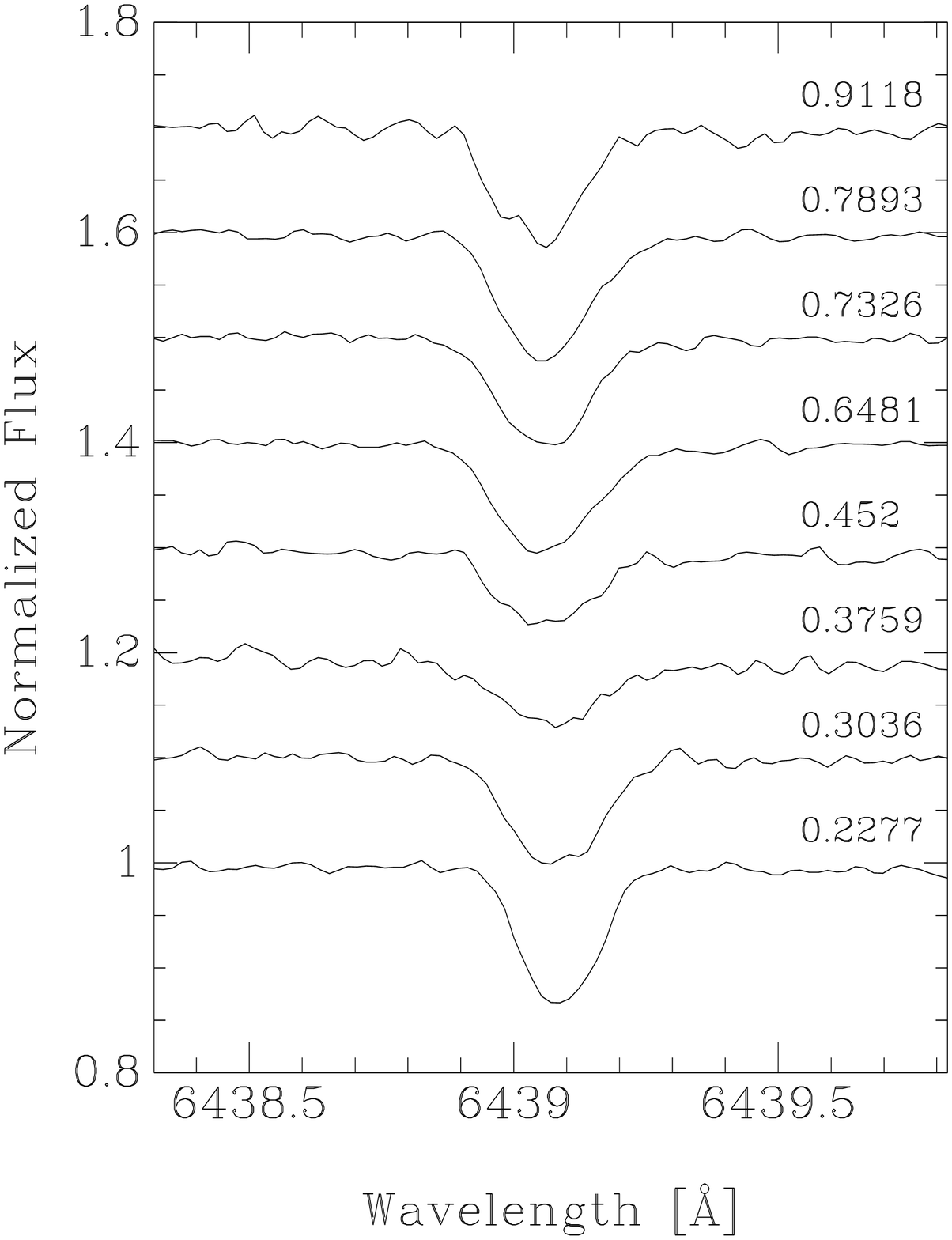}
\includegraphics[width=0.3\textwidth]{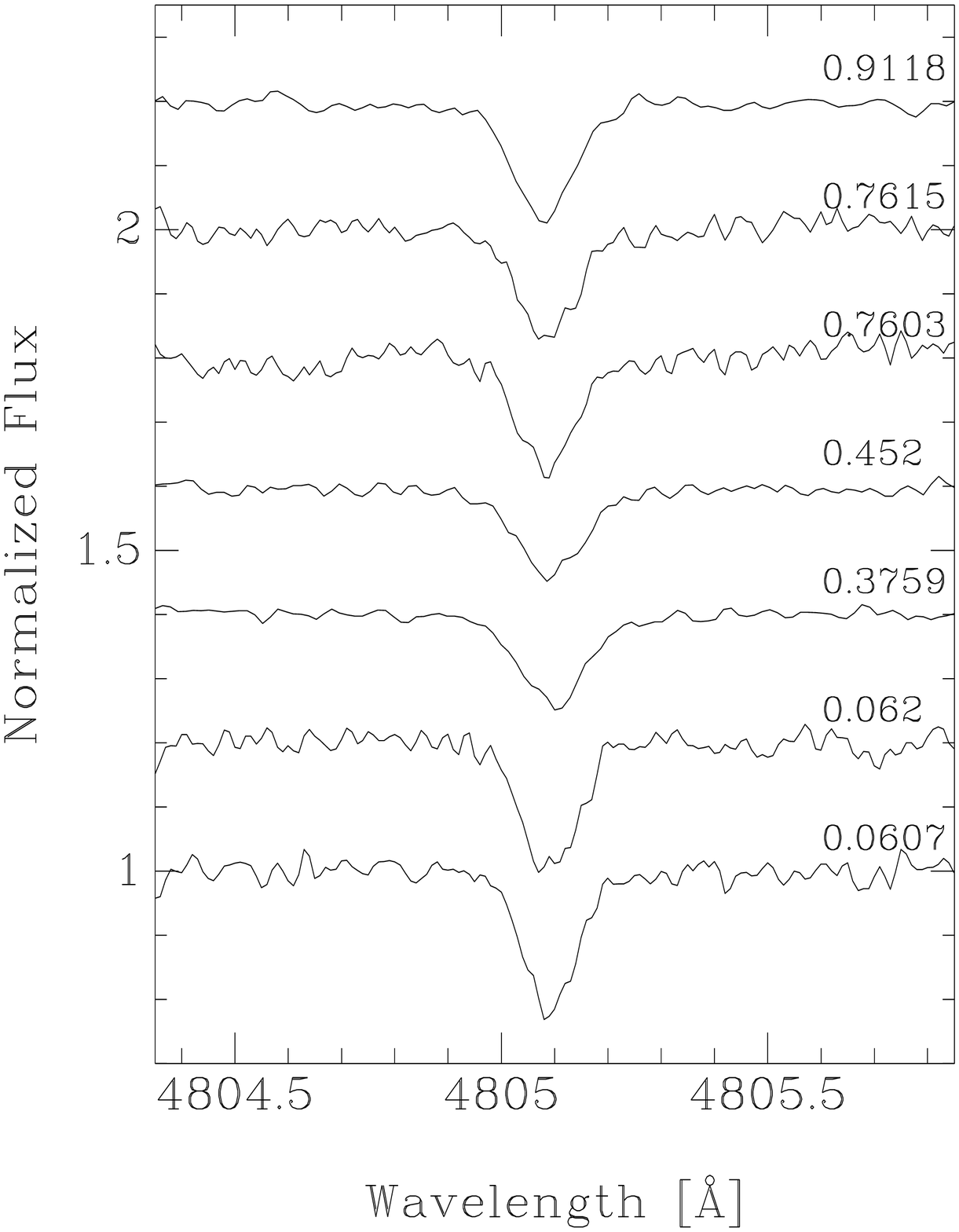}
\includegraphics[width=0.3\textwidth]{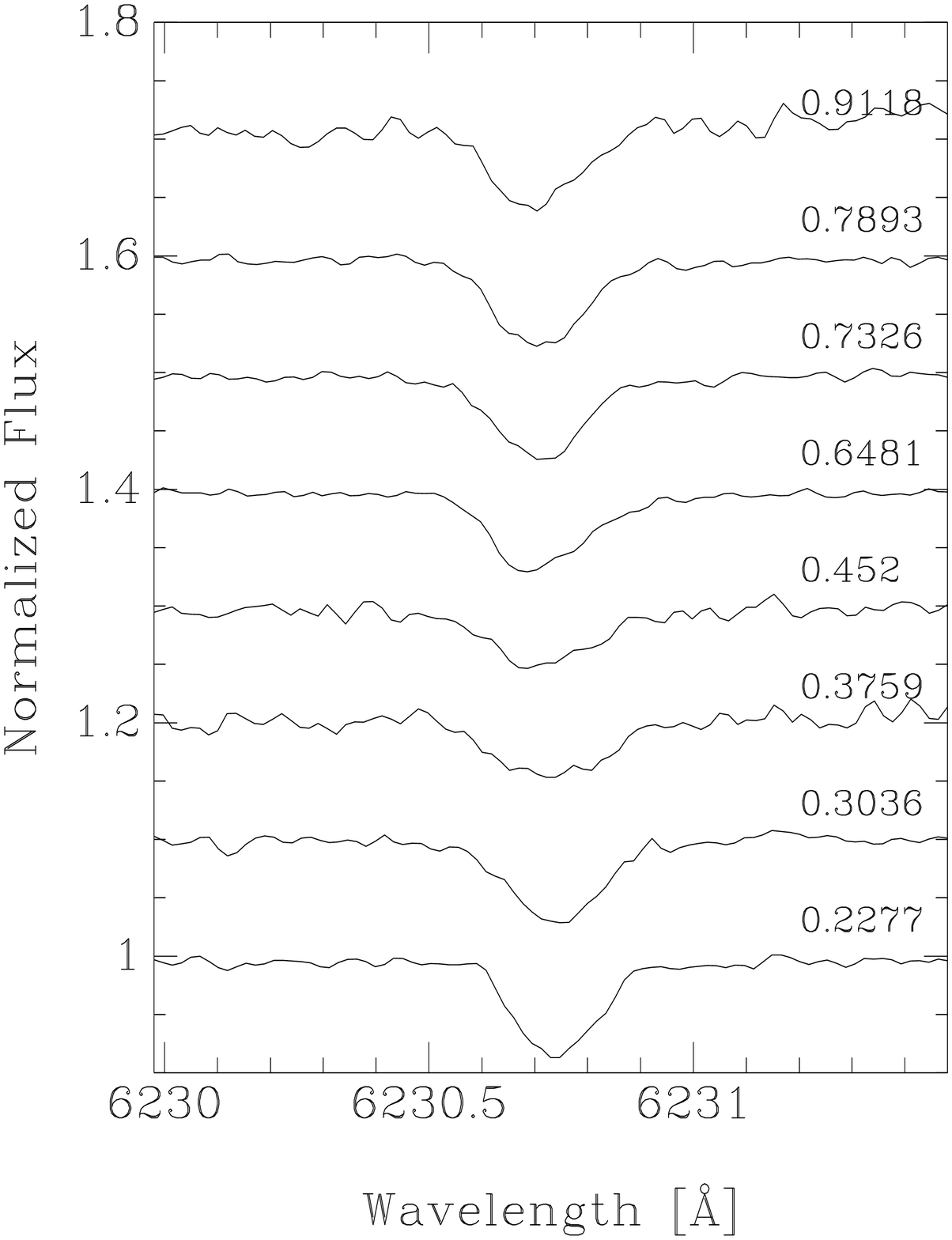}
\includegraphics[width=0.3\textwidth]{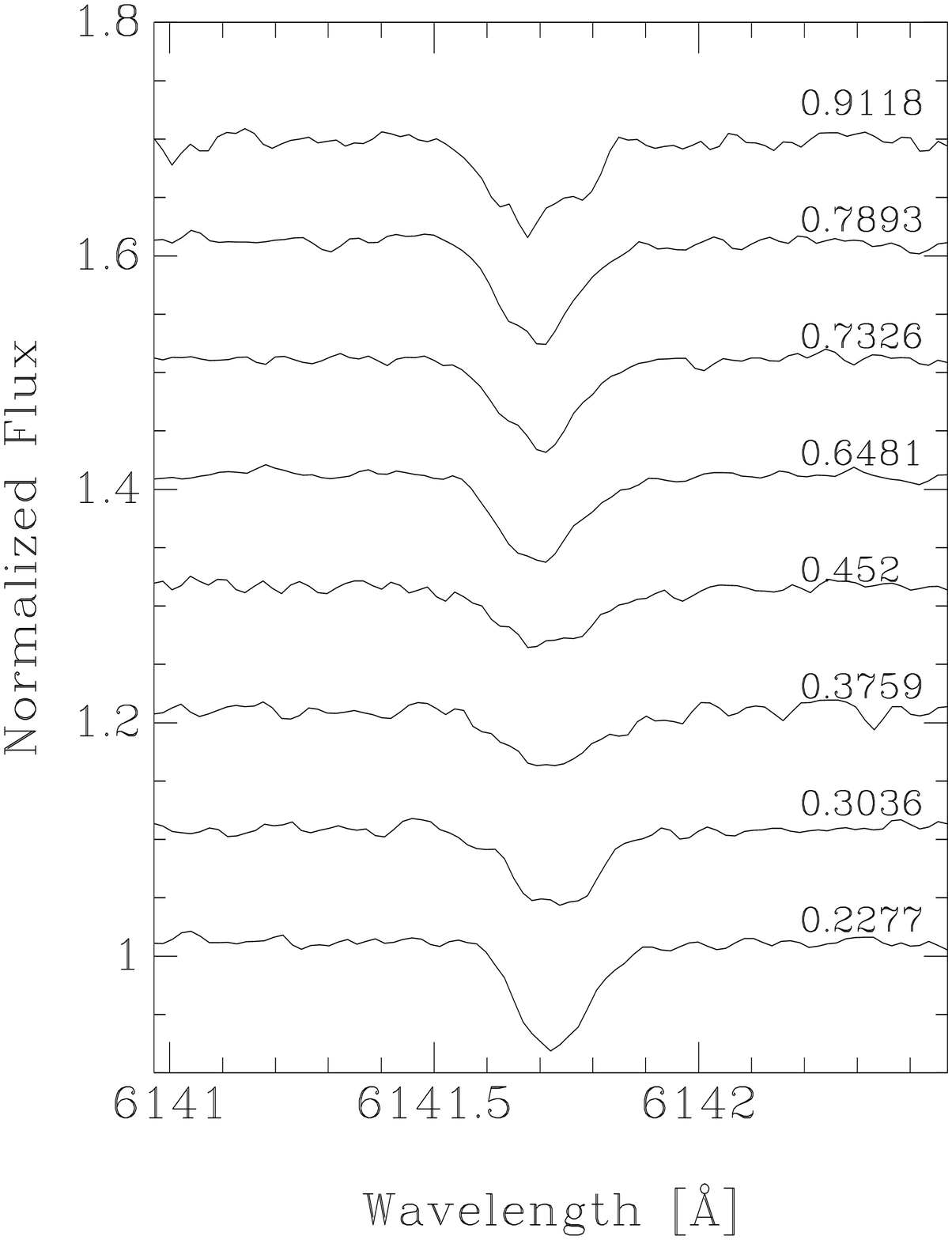}
\caption{
Variations of line profiles for different rotation phases in UVES spectra.
Upper panel, from left to right:
\ion{He}{i}~$\lambda$~6678.149,
\ion{Si}{ii}~$\lambda$~6347.091, and
\ion{Ca}{i}~$\lambda$~6439.073.
Lower panel, from left to right:
\ion{Ti}{ii}~$\lambda$~4805.105,
\ion{Fe}{ii}~$\lambda$~6230.856, and 
\ion{Ba}{ii}~$\lambda$~6141.718.
The rotation phase increases from bottom to top. 
Individual spectra are shifted in vertical direction for clarity.
}
\label{fig:var}
\end{figure*}

An example of the line profile variations for a few elements with this period is presented 
in Fig.~\ref{fig:var}. Our study reveals that the 
character of variability is
different for different elements, as it is usually expected for a horizontal inhomogeneous element distribution
over the stellar surface of Ap stars. 
Further, the profiles 
of \ion{Mg}{ii} $\lambda$~4481 and of the \ion{Ca}{ii} K lines exhibit strong broad wings and sharp cores,
which cannot be fitted with the same abundance, hinting at a vertical stratified abundance of these elements.
Such appearance and variations of metal line profiles are reminiscent of chemically peculiar 
main-sequence magnetic Ap stars, where the variability of chemical elements, especially of rare earth elements,
is one of the defining characteristics. 
On the other hand, no spectral lines belonging to exotic elements, such as
the lanthanide rare earths, or heavier elements were identified in our spectra.
We note that this is the first time the presence of element spots is detected on the surface of a Herbig Ae star. 

Preliminary indications from the \ion{Fe}{i}/\ion{Fe}{ii} equilibrium,
as well as Balmer profile fits suggest that the temperature is lower than
the values one might infer from the spectral type B9/A0V.
We estimate $T_{\rm eff}$ between 8000 and 9000\,K.  The low Balmer
lines (apart from H$\alpha$) are well fitted by a model with
$T_{\rm eff}$=8300\,K, log\,$g$=3.8,
but reasonable fits are also obtained with
$T_{\rm eff}$=8800\,K, log\,$g$=3.8.  If
the temperature is as low as 8300\,K, iron could be as much as a factor of
4 underabundant.

\section{Discussion}

Longitudinal magnetic fields of the order of a few hundred Gauss 
have  been detected in about a dozen Herbig Ae stars (e.g., Hubrig et al.\ \cite{Hubrig04}, 
\cite{Hubrig06}, \cite{Hubrig2007}, \cite{hubrig09};
Wade et al.\ \cite{Wade2005}, \cite{Wade2007}; Catala et al.\ \cite{Catala2007}).
For the majority of these stars rather small fields were measured, of the order of only 
100\,G or less. 
Our observations revealed that HD\,101412 possesses the strongest 
magnetic field ever measured in any Herbig Ae star, 
with a surface magnetic field $\left<B\right>$ up to 3.5\,kG. 
HD\,101412 is the first Herbig Ae star for which the rotational
Doppler effect was found to be small in comparison to the magnetic splitting 
and  several spectral lines
observed in unpolarized light at high dispersion are resolved into 
magnetically split components.
It is also the first time that the presence of element spots is detected on the 
surface of a Herbig Ae star.
The most pronounced variability was detected for spectral lines of 
\ion{He}{i} and the iron peak elements,
whereas the spectral lines of CNO elements are only slightly variable.
{Due to its very young age serves HD\,101412 as a unique object, setting the timescale for the development of 
photospheric chemical peculiarities.
Wade et al.\ (\cite{Wade2005}) claimed that another Herbig Ae star, HD\,72106A, shows abundance patches on it surface.
However, Folsom et al.\ (\cite{Folsom2008}) showed that this star is a bona fide young Bp star whereas the 
companion HD\,72106B is 
actually the Herbig Ae star displaying neither the presence of a magnetic field nor of element spots.}

{  Our previous study of Herbig Ae stars revealed a trend towards 
stronger magnetic fields for younger Herbig Ae stars, confirmed by statistical tests.
This is in 
contrast to a few other (non-statistical) studies claiming that magnetic Herbig Ae stars are progenitors of 
the magnetic Ap stars (e.g. Wade et al.\ \cite{Wade2005}). The disappearance of magnetic Herbig Ae stars 
(Hubrig et al.\ \ \cite{hubrig09})
and the emergence of Ap stars during the main-sequence life 
(e.g., Hubrig et al.\ \cite{Hubrig00}, \cite{Hubrig07}) are an indication that magnetic fields are not necessarily 
passive fossil fields leading to the observed features. Instead, 
the present observations of HD\,101412 are compatible with the theoretical scenario that 
the strong fields of Ap stars and magnetic fields in Herbig Ae stars are the result of 
a magnetic instability of internal toroidal fields. The current-driven 
Tayler instability (e.g., Vandakurov \cite{Vand72}, Tayler \cite{Tayler73}) is suppressed for 
very fast rotation. HD\,101412 is most likely a slower rotator, for 
which the Tayler instability could set in more easily, even during 
its pre-main-sequence phase. The instability delivers surface 
poloidal magnetic fields, which can be observed in contrast to the 
internal toroidal fields. At an estimated age of 2\,Myr, the star 
has passed its surface acceleration by accretion (St\c{e}pie\'n \cite{Ste00}). 
That exerts surface torques, while the interior is likely to be 
rotating differentially, thereby winding up strong enough toroidal 
magnetic fields to be supercritical for the Tayler instability.

A simplified estimate of the toroidal magnetic field strength necessary 
for the Tayler instability gives about 1~MG, according to the relation 
by Pitts \& Tayler (\cite{pit85}) saying that the Alfv\'en velocity should be 
smaller than the rotational velocity for stability. These fields are 
in principle possible to be created by differential rotation, but also 
smaller field strengths -- even below 100~kG in our example -- are 
unstable in a more sophisticated treatment, at the expense of considerably 
longer growth times (R\"udiger \& Kitchatinov \cite{rud10}). 
The growth time for a Pitts-and-Tayler field strength is only
a few rotations. When a 100-kG field becomes unstable, the
considerably longer growth time will still be of the order of 
years, which is much shorter than the evolutionary timescale.
The observed fields are non-axisymmetric remainders emerging
from the instability and are expected to be one order of magnitude 
weaker than the original, unstable toroidal field.

The rather strong magnetic field of the star HD\,101412 could thus be an example of an early emergence
of magnetic fields due to slower rotation or unknown environmental
differences compared to other Herbig Ae stars (Arlt \& R\"udiger, in preparation). Other A stars can suppress the
onset of the Tayler instability because of their faster rotation
and may turn into Ap stars only after their pre-main-sequence life, 
when winds have taken away a considerable amount of angular momentum.
While the mechanism of producing the surface poloidal fields would be 
the same, the observed magnetic Herbig Ae stars are not progenitors of Ap stars.
}

Generally, stellar magnetic fields in A-type stars on the main sequence are not symmetric relative to the 
rotation axis, so that the polarization 
signal is changing with the same period as the stellar rotation. The most simple modeling 
includes a magnetic field approximated by a dipole with the magnetic axis inclined to the rotation axis.
The vast majority of the studied Ap stars with magnetically resolved lines have average magnetic field moduli in 
the range 3--9\,kG. It was previously discussed by Mathys et al.\ (\cite{mathys97}) that 
the low field end shows a rather sharp cutoff. 
Indeed, the magnetic field moduli extends below 3kG only for two stars,
reaching 2.9\,kG (HD\,29578) and 2.7\,kG (HD\,75445), although we would expect to be
able to detect resolved magnetically split lines for weaker fields,
down to 1.7\,kG. 
For HD\,101412, the average of the mean field 
modulus is  2.6\,kG, determined from UVES spectra with highest S/N ratios. This value is less than the threshold of 2.7\,kG, 
but still much larger than 1.7\,kG.
Thus, we confirm the previous conclusion of Mathys et al.\ (\cite{mathys97}) that the absence
of any star with a phase-averaged field modulus lower than 2.5--2.7\,kG
among the known stars with resolved magnetically split lines
is not due to observational limitations, but rather reflects an intrinsic stellar property.
This result is puzzling as, similar to Ap stars, 
the distribution of the mean longitudinal field in Herbig Ae stars is strongly skewed towards 
small field values, down to the 
limit of detectability (e.g.\ Hubrig et al.\ \cite{hubrig09}).  
Since the age of HD\,101412 is only $\sim$2\,Myr,
this star is observed at an evolutionary stage where the magnetic field plays a critical role in controlling 
accretion and stellar wind.  
Due to the insufficient number of magnetic field measurements, in particular of mean longitudinal 
magnetic field measurements, which are sensitive to field geometry,  the real structure of the magnetic field 
remains presently unknown.
Clearly, the strong magnetic nature of 
this object makes it a prime candidate for future studies of the relation between magnetic field and physical 
processes occurring during stellar formation.



\label{lastpage}

\end{document}